# Reciprocal Symmetry and Unified Classico-Quantum Oscillator

# And

# Consistency between a Particle in a Potential Well and a Harmonic Oscillator

Mushfiq Ahmad

Department of Physics, Rajshahi University, Rajshahi, Bangladesh

E-mail: mushfiqahmad@ru.ac.bd

Muhammad O. G. Talukder

Department of ETE and CSE, University of Development Alternative, Dhaka, Bangladesh

**Abstract**

The function exp(iwt) describes an oscillating motion. Energy of the oscillator is proportional to the square of w. exp(iwt) is the solution of a differential equation. We have replaced this differential equation by the corresponding finite-time difference equation with d as the smallest span of time. We have, then, symmetrized the equation so that it remains invariant under the change d going to -d. This symmetric equation has solutions in pairs. The angular speed w is modified to w' or w". w' contains a part with an integer. w" contains a part with a half-integer. This corresponds to quantum mechanical oscillator energy levels. F= a.exp(iwt) describes oscillation between -a and +a. If we make w=0, F describes free oscillation between -a and +a. Reciprocal symmetric oscillator, thus, unifies quantum and classical harmonic oscillators on one hand, and a harmonic oscillator and a free particle in a potential well on the other hand. No quantum mechanical postulates are involved.

## 1. Introduction

The differential equation

$$\frac{df}{dt} = iwf \qquad (1.1)$$

has the solution, $f = a.\exp(iwt)$. $f$ describes the motion of a harmonic oscillator if

$$w = \sqrt{\frac{k}{m}} \qquad (1.2)$$

Where $k$ is spring constant and $m$ is the mass.

Classical energy of the oscillator $E_{classical}$ is proportional to $w^2$

$$E_{classical} = \frac{1}{2}m(a.w)^2 \qquad (1.3)$$

The corresponding finite difference equation has solutions[1], which come in reciprocal pairs[2]. When the function represents a harmonic oscillator, both the solutions will contribute to oscillator energy. We intend to study these contributions and compare them to corresponding classical and quantum mechanical values.

Let the oscillator having amplitude $a$ oscillate between $-a$ and $+a$. We place two perfectly rigid reflecting walls at $-a$ and $+a$. The presence of the walls does not influence the oscillation in any way. Now we make $k = 0$ so that $w = 0$. This makes the oscillator a free particle. Therefore, in this case ($w=0$) the non-vanishing energy term should the energy levels of a free particle bouncing between reflecting walls.

We intend to study the energy levels of a reciprocal symmetric harmonic oscillator, compare them with the corresponding quantum mechanical energy levels, and also study its behavior in the limit $w \to 0$.

## 2. Oscillator Finite Difference Equation

To exploit the symmetry properties of a harmonic oscillator we replace (1.1) by the corresponding symmetric finite difference equation[3]

$$\frac{Dg_\pm}{D(t,d)} = iw.g_\pm \qquad (2.1)$$

where

$$\frac{Dg_\pm(w,t)}{D(t,d)} = \frac{g_\pm(w,t+d) - g_\pm(w,t-d)}{2d} \qquad (2.2)$$

The above difference quotient has the following symmetry under the change $\delta \to -\delta$

$$\frac{Dg_{\pm}}{D(t,-d)} = \frac{Dg_{\pm}}{D(t,d)} \tag{2.3}$$

We require that (2.1) should go over to (1.1) in the limit $d \to 0$

$$\frac{Dg_{\pm}}{D(t,d)} = iwg_{\pm} \xrightarrow{d \to 0} \frac{df}{dt} = iwf \tag{2.4}$$

(2.1) has solutions in pairs. One of the pairs is[4]

$$g_{+} = \exp\left(\frac{2\pi}{d}s_{+} + \frac{\sin^{-1}(wd)}{d}\right)it = \exp(w_{+}it) \tag{2.5}$$

$$g_{-} = \exp\left(\frac{2\pi}{d}s_{-} - \frac{\sin^{-1}(wd)}{d}\right)it = \exp(w_{-}it) \tag{2.6}$$

where

$$s_{+} = \text{integer} \tag{2.7}$$

$$s_{-} = \text{half - integer} \tag{2.8}$$

With the correspondence relation (for $s_{+} = 0$)

$$g_{+} \xrightarrow{d \to 0} f \tag{2.9}$$

$g_{+}$ and $g_{-}$ are related through the reciprocity relation

$$g_{+} \cdot g_{-} = (-1)^{t/d} \tag{2.10}$$

## 3. Classical and Quantum Energy Levels

The energies of the reciprocal symmetric oscillator $E_{\pm}^{R.S}$ are

$$E_{+}^{R.S} = \frac{1}{2}m(a.w_{+})^2 = \frac{1}{2}m.a^2[\{(2\pi s_{+})/d\}^2 + 2\{(2\pi s_{+})/d\}w + w^2] \tag{3.1}$$

$$E_{-}^{R.S} = \frac{1}{2}m(a.w_{-})^2 = \frac{1}{2}m.a^2[\{(2\pi s_{-})/d\}^2 - 2\{(2\pi s_{-})/d\}w + w^2] \tag{3.2}$$

For $s_{+} = 0$ (3.1) gives the classical value (1.3).

The middle term of (3.2) is

$$E_{-\text{middle term}}^{R.S} = -m.a^2(\pi/d)(2s_{-})w \tag{3.3}$$

It corresponds to quantum mechanical value[5].

$$E_{quantel}^{oscillator} = \eta(2s_-)w \tag{3.4}$$

The important difference is that there is no Planck's constant in (3.3).

## 4. Square Well Potential Energy Levels

Let the oscillator having the amplitude $a$ oscillate between $-a$ and $+a$. There are two perfectly rigid reflecting walls placed at $-a$ and $+a$. The presence of the walls does not influence the oscillation in any way. Now we make $w=0$ so that the oscillator becomes a free bouncing particle. Of the energies (3.1) and (3.2) we are left with the first parts only

$$E_+^{R.S} = \frac{1}{2}m(a.w_+)^2 = \frac{1}{2}m.(a/d)^2(\pi.2s_+)^2 \tag{4.1}$$

$$E_-^{R.S} = \frac{1}{2}m(a.w_-)^2 = \frac{1}{2}m.(a/d)^2(\pi.2s_-)^2 \tag{4.2}$$

Combining (4.1) and (4.2) we have

$$(E_+^{R.S} + E_-^{R.S}) = \frac{1}{2}m(a/d)^2\pi^2\{(2s_+)^2 + (2s_-)^2\} = \frac{1}{2}m(a/d)^2(\pi.2s)^2 \tag{4.3}$$

Where $2s$ is an integer.

Compare (4.3) with $E_{quantel}^{s.w.}$, the quantum mechanical energy levels of one dimensional square well potential[6]

$$E_{quantel}^{s.w.} = \frac{1}{2m}\left(\frac{\eta}{2a}\right)^2 \pi^2\{(2s_+)^2 + (2s_-)^2\} = \frac{1}{2m}\left(\frac{\eta}{2a}\right)^2 (\pi.2s)^2 \tag{4.4}$$

Apart from the constants and the dependences on $d$ and $\eta$ (4.3) and (4.4) compare well.

## 5. Assumption of Classical Physics

Consider an oscillator oscillating along $x$ line between $-a$ and $+a$

$$x = a.\sin wt \tag{5.1}$$

$x = 0$ at $t=0$. We measure at intervals of $d$. We are not able to measure at any interval less than $d$. After time $d$ the value of $x$ is

$$x = a.\sin wd \tag{5.2}$$

What is the value $x$ after time $d'$, where $d' < d$ ? The classical *assumption* is

$$x = a.\sin wd' \tag{5.3}$$

It is an assumption because no observations have been made for any interval $d' < d$. In (2.9) and (2.10) of this paper we have replaced assumption (5.3) by the less stringent assumptions below and we write for the displacement $x$

$$x_+ = a.\sin(\frac{2\pi}{d}s_+ + w)it = a.\sin(w_+ it) \tag{5.4}$$

$$x_- = a.\sin(\frac{2\pi}{d}s_- - w)it = a.\sin(w_- it) \tag{5.5}$$

We require that $x_+$ and $x_-$ agree with the observed value at $t = d$ so that

$$x_+ = a.\sin(w_+ id) = x_- = a.\sin(w_- id) = x = a.\sin(wid) \tag{5.6}$$

(5.4) and (5.5), therefore, express our ignorance about the values of $x$ for $t < d$.

## 6. Conclusion

The pair of reciprocal symmetric functions $g_\pm$ of (2.9) and (2.10) describes a classical oscillator, which has discrete energy levels (3.1) and (3.2). One of the terms (3.3) compares well with the corresponding quantum mechanical term (3.4).

If we set $w=0$, $g_\pm$ give the energy levels (4.3) of a free particle bouncing between reflecting walls. These levels correspond to quantum mechanical one-dimensional potential well levels (4.4). Reciprocal symmetric oscillator, thus, unifies both classical and quantum mechanical features.

Planck's radiation law goes over to corresponding classical Rayleigh-Jeans[7] law as $\eta \to 0$.

$$\frac{\eta w}{\exp(\eta w/kT) - 1} \xrightarrow{\eta \to 0} kT \tag{6.1}$$

One would desire that, similarly, quantum mechanical oscillator energy (3.4) should go over to corresponding classical oscillator energy (1.3) as $\eta \to 0$. This does not happen.

This shows a lack of correspondence between classical mechanics and quantum mechanics.

In the limit $w \to 0$ the oscillator becomes a free particle oscillating between $-a$ and $+a$. This is the same as a free particle oscillating between rigid walls. This is equivalent to a potential well. Therefore, in this limit the oscillator energy levels (3.4) should go over to the corresponding potential well energy levels (4.4). This does not happen. This shows a lack of correspondence within quantum mechanics itself.

Reciprocal symmetric oscillator (described by (3.1) and (3.2)) incorporates (1.3), (3.3) and (4.2).

---

[1] Mushfiq Ahmad. Reciprocal Symmetry and Equivalence between Relativistic and Quantum Mechanical Concepts. http://www.arxiv.org/abs/math-ph/0611024

[2] Mushfiq Ahmad. Reciprocal Symmetric Boltzmann Function and Unified Boson-Fermion Statistics. Not published

[3] Mushfiq Ahmad. Reciprocal Symmetric and Origin of Quantum Statistics. http://www.arxiv.org/abs/physics/0703194

[4] M. Osman Gani Talukder. An Alternative Approach to The Relativity. Grontho Procation. 11 New Market, Rajshahi-6100, Bangladesh

[5] Introduction to Quantum Mechanics. Robert H. Dicke and James P. Wittke. Addison-Wesley Pub. Co., Inc.

[6] Introduction to Quantum Mechanics. Robert H. Dicke and James P. Wittke. Addison-Wesley Pub. Co., Inc.

[7] http://en.wikipedia.org/wiki/Rayleigh-Jeans_law